\documentclass[10pt,letterpaper]{article}
\usepackage{template}
\usepackage{cite}
\usepackage[usenames,dvipsnames,svgnames,table]{xcolor}

\begin{document}

\title{Bragg-Scattering conversion at telecom wavelengths  \textcolor{Black}{towards} the photon counting regime }

\author{Katarzyna Krupa,$^{1,*}$ Alessandro Tonello,$^{1}$ Victor V. Kozlov,$^{2}$ Vincent Couderc,$^{1}$ 
Philippe Di Bin,$^{1}$ Stefan Wabnitz,$^{3}$ Alain Barth\'el\'emy,$^{1}$ Laurent Labont\'e$^{4}$ and S\'ebastien Tanzilli$^{4}$}

\address{}
\address{1) Universit\'e de Limoges, XLIM, UMR CNRS 7252, \\ 123 Av. A. Thomas,  87060 Limoges, France\\
2) Department of Physics, St.-Petersburg State University, Petrodvoretz, \\ St.-Petersburg, 198504, Russia\\
3) Dipartimento di Ingegneria dell'Informazione, Universit\`a di Brescia, \\ via Branze 38, 25123 Brescia, Italy\\
4) Universit\'e de Nice-Sophia Antipolis, Laboratoire de Physique de la Mati\`ere Condens\'ee (LPMC), UMR CNRS 7336, \\ Parc Valrose, 06108 Nice, France}

\email{* katarzyna.krupa@xlim.fr} 



\begin{abstract}
\textcolor{Black}{We experimentally study Bragg-scattering four-wave mixing in a highly nonlinear fiber at telecom wavelengths using photon counters. We explore the polarization dependence of this process with a continuous wave signal in the macroscopic and attenuated regime, with a wavelength shift of 23 nm. Our measurements of mean photon numbers per second under various pump polarization configurations agree well with the theoretical and numerical predictions based on classical models. We discuss the impact of noise under these different polarization configurations. }
\end{abstract}





\section{Introduction}
Current studies of four-wave mixing (FWM) in third-order nonlinear media, such as hot atomic vapours, silicon waveguides or silica fibers, aim to develop high-performance devices for ultrafast, all optical, signal data processing \cite{Kazo,Ding1,Alic2}. In the context of quantum computing and communication, Bragg Scattering type FWM (BS-FWM) or frequency exchange \cite{Kazo} has a high potential for achieving coherent single photon manipulation and entanglement-preserving frequency conversion, as it has already been developed in second-order nonlinear crystals \cite{Langford3, Tanz, Zei, Noe, NTT}. The BS-FWM can actually provide a simple, efficient, and noiseless means for photon-photon optical frequency conversion. This has been recently demonstrated both theoretically \cite{McKinstrie4} and experimentally \cite{McGuinness5} taking advantage of special photonic crystal fibers operating in the visible range of wavelengths. A very recent work shows low-noise BS-FWM in $SiN_x$ waveguides, with pumps at 1550 nm and signals and idler at 980 nm \cite{Si}. In view of possible integration in single photon based quantum communication fiber networks, extending BS-FWM based frequency converters to the telecom range for signal and idler is of high interest. \textcolor{Black}{In another recent work coherent FWM has been proposed for shifting on-demand the carrier frequency of single photons at telecom wavelengths in the pulsed regime \cite{Clemmen}. }

	In this work, for the first time to the best of our knowledge, we investigate experimentally the \textcolor{Black}{ polarization properties of} frequency conversion of an attenuated continuous-wave (CW) telecom laser using vector BS-FWM \textcolor{Black}{and CW pumps}. That is, a signal field at 1549.2 nm is converted into an idler field at 1526.43 nm. \textcolor{Black}{For the first time we also discuss some of the limiting factors on exploitability of BS-FWM coming from noise. In our setup it turns out that the noise follows the spectral distribution of Raman gain: we study its polarization dependence as this represents one of the crucial issues towards the implementation at the single photon regime.} Because of the anisotropic nature of the Raman gain, we show that both parallel and perpendicular contributions to the Raman noise are significant for relatively small frequency detunings \cite{Li6}.

\section{Experimental setup}
\begin{figure}[ht]
 \centering
 \includegraphics[height=41mm]{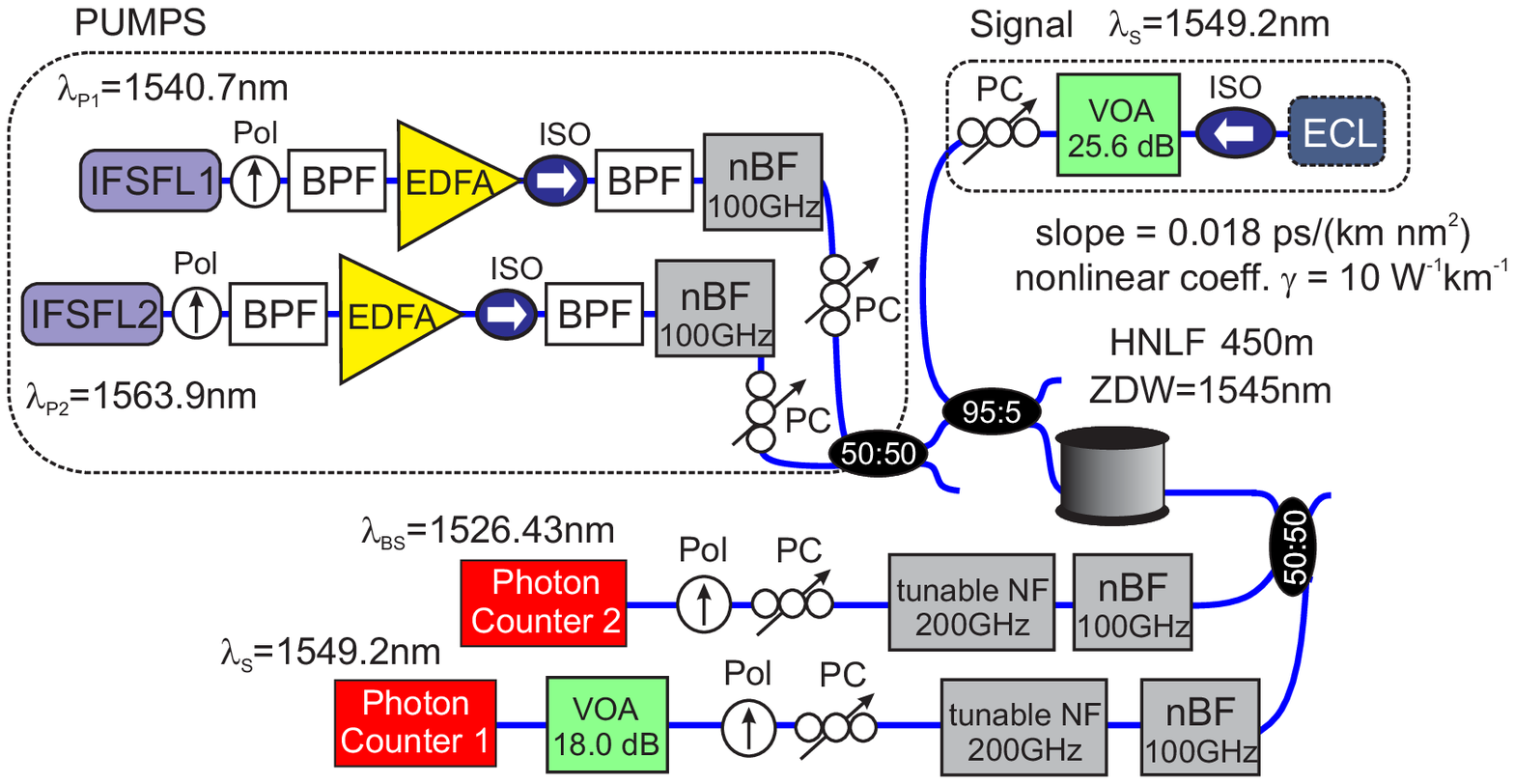}
\caption{\label{Fig1} Experimental setup: See the text for details.}
\end{figure}

Our all-fiber experimental setup is shown in Fig.\ref{Fig1}. To mitigate Brillouin scattering in the fiber, we took advantage of two independent linearly polarized intra cavity frequency-shifted feedback laser (IFSFL) pumps \cite{Krupa7} operating in the CW regime at 1540.7 nm (P1) and 1563.9 nm (P2). The two IFSFLs were amplified separately using two erbium-doped fiber amplifiers (EDFA), which were filtered and isolated from unwanted back reflections. The ASE noise from the amplifiers was reduced thanks to inline thin-film bandpass filters (BPF). Next, to further reduce the background spectral level, each pump was filtered using a 100 GHz wideband filter (nBF) having a noise rejection of about 60 dB (AOS GmbH). The signal (S) at 1549.2 nm was provided by a 5 mW extended cavity laser (ECL) attenuated by 25.6 dB when working with photon counters. We controlled the polarization state of the two pumps and the signal by means of fiber polarization controllers (PC). The two pumps were combined at a 50/50 fiber coupler, while the signal was added through a 5/95 ÒtapÓ coupler, which provided a further insertion loss of 13 dB for the signal channel. We used a $L=450 m$ long highly nonlinear fiber (HNLF) whose zero dispersion wavelength (ZDW) was at 1545 nm. \textcolor{Black}{ The fiber dispersion slope was  $0.018 ps/(km\cdot nm^2)$ and the fiber nonlinear coefficient was of $\gamma=10 W^{-1}km^{-1}$ (see also Ref. \cite{180nm}.} 
\textcolor{Black}{At the HNLF input the average powers of pumps were 22 mW (P1) and 15 mW (P2). The spectral bandwidth for pumps were of 44 GHz for P1 and 50 GHz for P2 (see Ref. \cite{Krupa7} for details). The ECL laser bandwidth was of 100kHz.}

At the fiber output, we placed a second 50/50 coupler to analyze two spectral regions simultaneously: one at 1549.2 nm (the residual signal) and one at 1526.43 nm (the idler, whose frequency is the sum of the signal frequency and the frequency difference between the two pumps).  We isolate signal and idler from the two intense pumps by an appropriate narrow filtering stage. Therefore, at each output arm of the coupler, we cascaded two filters (nBF filter followed by a Newport tunable filter) with a total noise rejection of about 100 dB. To explore different pump and signal polarization arrangements, we combined, at the output, the PC with a fiber polarizer (Pol). Depending on the intensity of the signal channel, we employed two different detection schemes. With a non attenuated signal, we directly measured the fiber output spectrum with an optical spectrum analyzer. On the contrary, when the signal was attenuated, we used two photon counters (IDQ 201) operating at 100 kHz trigger rate and 10\% detection efficiency.\textcolor{Black}{ We used a gate duration of 2.5 ns, and under these conditions the photon counters have a dark-count probability of  $2.7\times10^{-6} ns^{-1}$.}

\section{Frequency conversion with non attenuated signal: numerical simulations and experimental results}
\begin{figure}[ht]
 \centering
 \includegraphics[height=40mm]{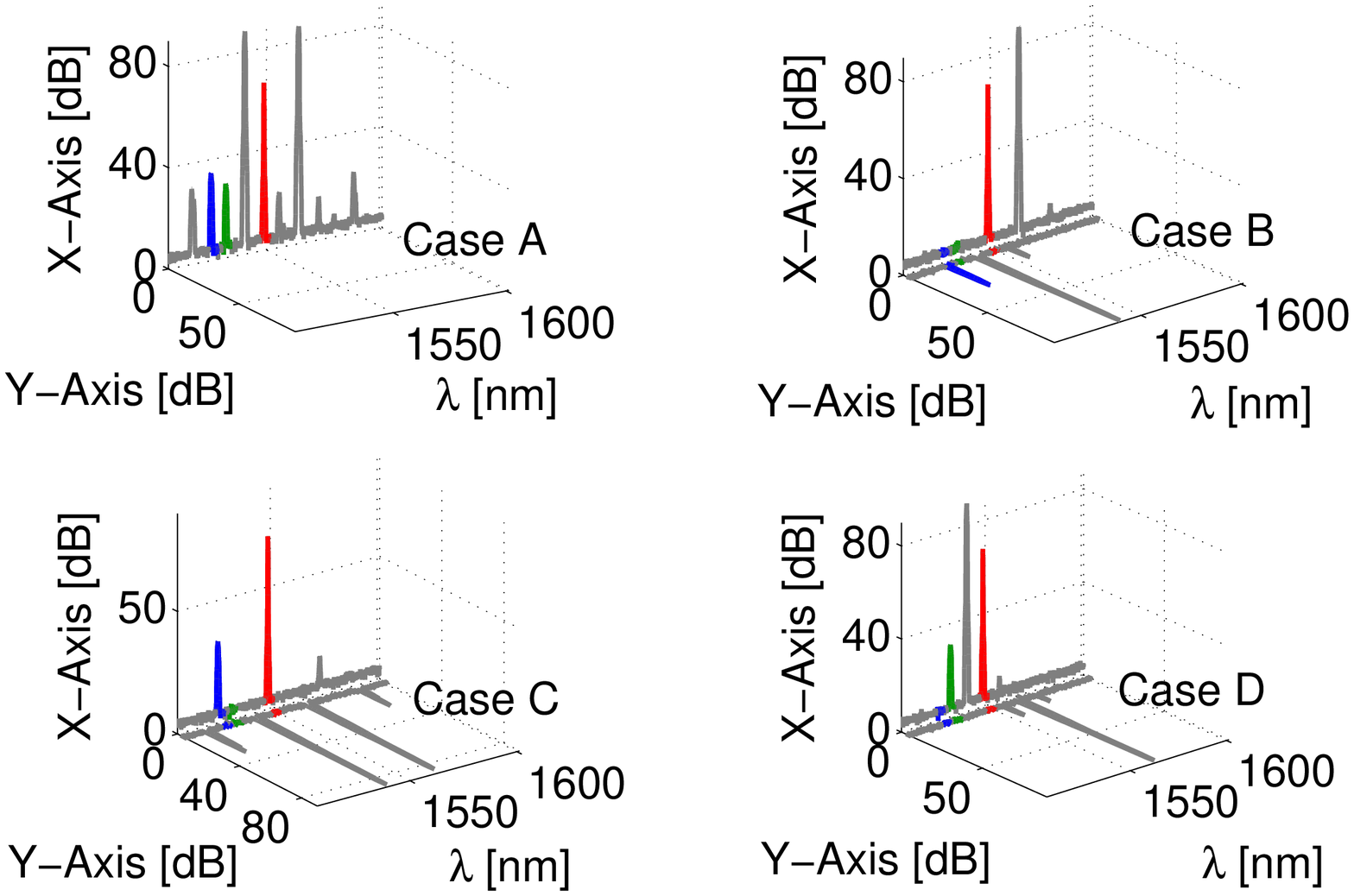}
 \includegraphics[height=35mm]{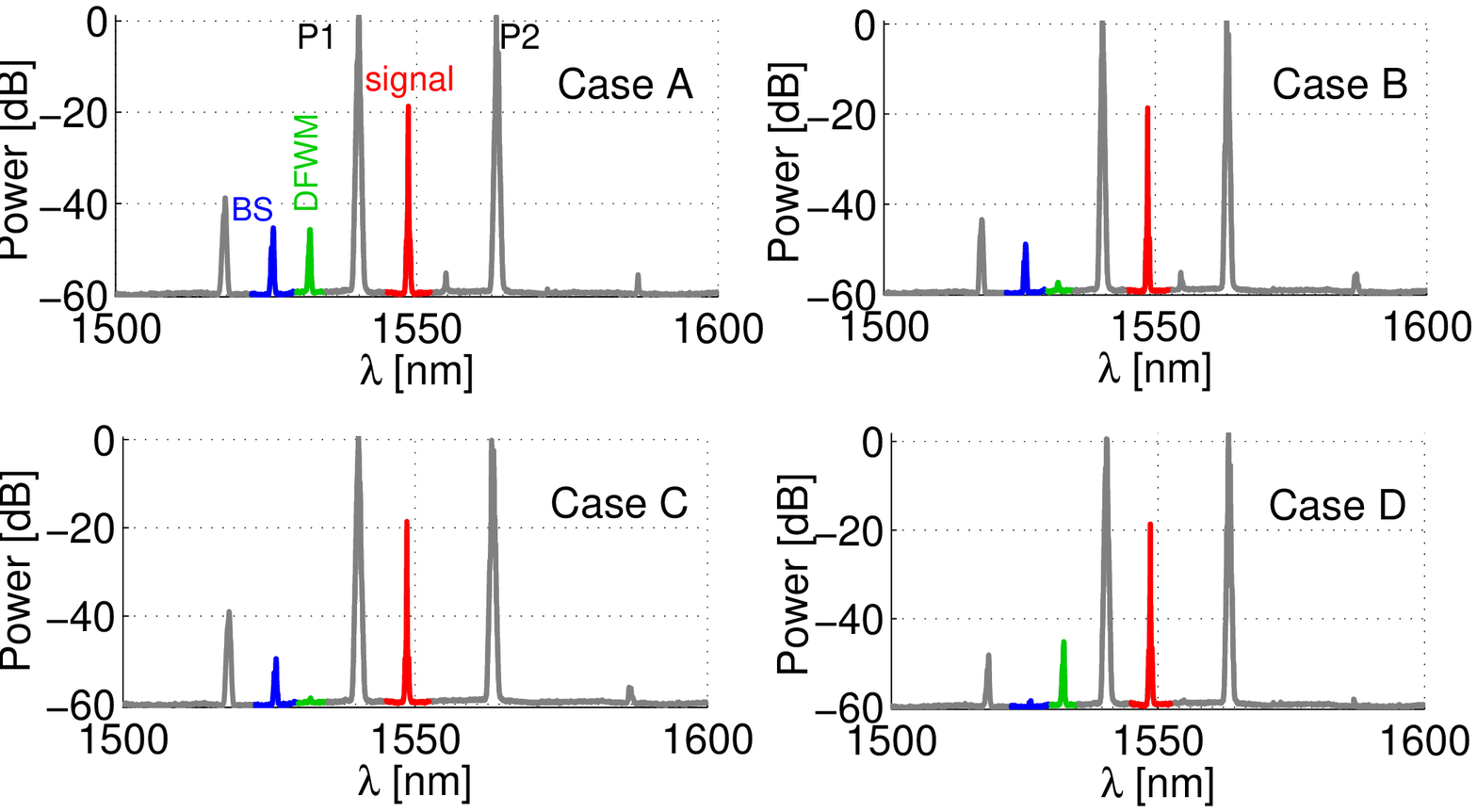} 
\caption{\label{Fig2} Left. Numerical simulations of dual pump FWM. Red: signal at 1549.2 nm;  Blue: BS idler at 1526.43 nm; Green: DFWM at 1532.3 nm. Cases A - D are described in the text.   Right. Experimental spectra from an optical spectrum analyzer in the strong signal regime. Cases A-D correspond to the numerical simulations shown of the left panel.}
\end{figure}

Following Ref. \cite{McKinstrie4}, we show first in the left panel of Fig. \ref{Fig2} the results of numerical simulations based on two incoherently coupled nonlinear Schr\"odinger equations (Manakov system) with parameters close to our experimental setup. We considered four different input conditions representing different combinations of polarization states for the pumps and the signal. \textcolor{Black}{The pump powers were set to a fixed values when comparing the effect of the different input polarization states in the numerical simulations and in the experiments}. To account for the broadband nature of the pumps, each plot represents the average spectral power taken over an ensemble of 20 numerical runs \cite{Krupa7}. In case A, pumps and signal are all co-polarized, leading to the strongest signal-idler coupling, hence the highest BS FWM conversion efficiency \cite{McKinstrie4}. \textcolor{Black}{In our implementation the low value of dispersion limits the conversion efficiency of BS-FWM (from 1549.2 nm to 1526.43 nm); moreover, other nonlinear wave mixing effects are also active.} For example, degenerate four-wave mixing (DFWM) involving P1 converts the same signal into an idler at 1532.3 nm. The relatively large intensity of the DFWM-induced idler is due to the proximity of the signal wavelength to the fiber ZDW. Additionally, spontaneous FWM induced by the two pumps only generates the two external sidebands in Fig.\ref{Fig2} left.
	In cases B and D, the two pumps are cross-polarized, so that only one of them is co-polarized with the signal. Instead in case C, the two pumps are co-polarized, while being cross-polarized with the signal and the BS-idler. Note that in cases B and C the conversion efficiency is 6 dB lower than in case A \cite{McKinstrie4}. However, DFWM is switched off due to the lack of parity in the number of photons per polarization state. The opposite happens in case D: here the signal is co polarized with P1, such that DFWM is active and BS is switched off, as predicted in Ref. \cite{McKinstrie4}. These numerical studies are in excellent agreement with the obtained experimental spectra shown in the right panel of the same Fig.\ref{Fig2}. More specifically, we observe the cases where both BS-FWM and DFWM are present (case A of Fig.\ref{Fig2}, right panel), and where they are mutually exclusive (cases B, C and D in Fig.\ref{Fig2}, right panel), in full agreement with the relevant polarization arrangements. Case C is of particular interest for quantum frequency conversion due to a reduced Raman noise \cite{Langford3,McKinstrie4}.

\section{Experimental results \textcolor{Black}{with an attenuated signal}}
\begin{figure}[ht]
 \centering
 \includegraphics[height=40mm]{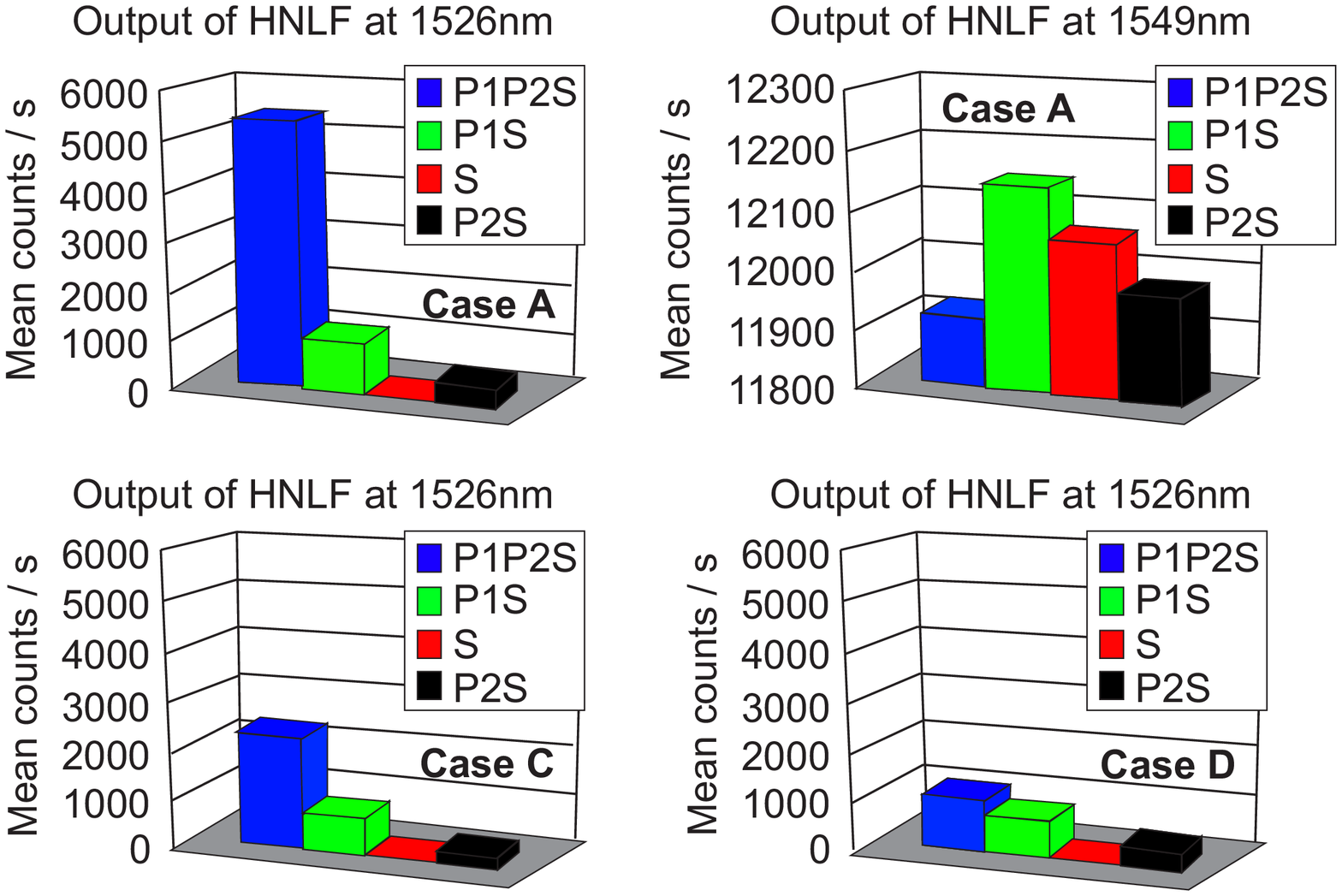}
 \includegraphics[height=35mm]{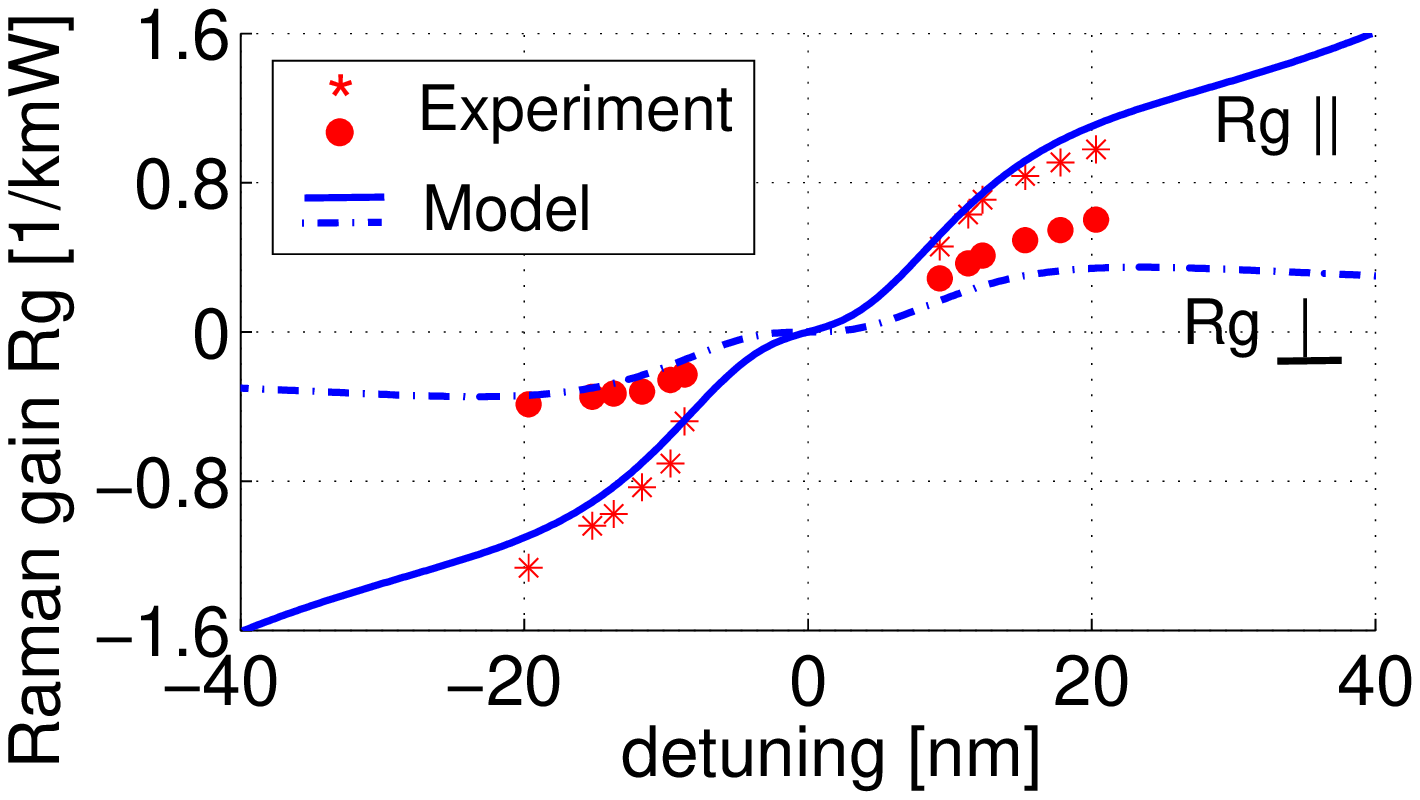} 
\caption{\label{Fig3} Left. Experimental results with photon counters (see text for details). Right. Raman gain calculated from noise measurements of an extended cavity laser (ECL) at 1540.7 nm.}
\end{figure}

\textcolor{Black}{ Let us now consider the problems arising when aiming at operating in the photon counting regime. Because of the low conversion efficiency of our experimental configuration, the simplest way to distinguish clearly the idler from noise, in all polarization configurations, was to provide a large enough signal power. For this reason we measured the residual signal after a 18dB attenuator (see Fig.\ref{Fig1}). The 3dB coupler and a chain of filters at the fiber output gave an insertion loss of 9dB. At the fiber output we estimated then 600 signal photons/gate. We underline that our signal remains an attenuated laser and not a single photon source. Note however that no attenuation was employed for the idler detection for which we estimate 5 photons/gate (the idler optical bandwidth at macroscopic signal was 75 GHz thanks to the IFSFLs). The experimental results are illustrated in the left panel of  Fig.\ref{Fig3}}. The histograms provide the mean number of detected photons per second, either at 1526.43 nm (BS-idler channel, see top left and bottom left and right panels), or at 1549.2 nm (signal channel, top right panel). \textcolor{Black}{We observe that the ratio between signal and idler counts has the same order of magnitude which was previously observed at the macroscopic level, when we consider the effect of the 18dB attenuator for the signal channel (see Fig.\ref{Fig2}).} For each case, the counts were recorded for four different input conditions. That is, when all P1, P2 and S are present (blue bars), when only P1 and S are present (green bars), when only S is present (red bars), and finally when only P2 and S are present (black bars). Dark counts were negligible being in the order of 1.2 counts/s. By looking at the top right panel, which corresponds to case A, we observe that the signal is slightly amplified by P1 (compare red and green bars) and slightly attenuated by P2 (compare red and black bars). Indeed, since the signal wavelength is in-between those of P1 and P2, the observed signal gain/loss is likely due to stimulated Raman scattering. When both pumps and the signal are present, a further signal depletion was measured (compare red and blue bars). Correspondingly, we observed a clear count enhancement at the BS-idler wavelength (see the blue bar of the top left histogram). Nevertheless, nonzero counts were observed at 1526.43 nm even when one of the pumps was switched off. We believe that this noise contribution is mainly induced by spontaneous Raman emission from the pumps, as will be discussed later on. Case A maximizes the idler counts, thanks to the associated larger BS-FWM conversion. This case also leads to the largest (parallel) Raman noise. By suitably choosing the polarization states of pumps and signal, we could vary the number of counts for both the BS idler (blue bars) and the Raman noise (green and black bars) in the 1526 nm channel. For instance, in case D, we could reduce the BS counts down to the noise level, despite the presence of both pumps and the signal. This is due to the switching off of the BS-FWM (see Fig.\ref{Fig2}). Here, the blue bar combines the Raman noise carried by the two pumps, and its level is equal to the sum of the green and black bars. Finally, in case C we could reach an intermediate level of conversion: in this situation the sidebands experience perpendicular Raman gain from the pumps. Note that the overall performances of cases A and C are similar, namely in terms of nearly equal signal-to-noise ratio for the BS idler. With respect to case A, in case C we observe a reduction of the Raman noise (which still remains nonzero due to the perpendicular Raman contribution, see the right panel of Fig.\ref{Fig3}) simultaneously with a nearly equal reduction of the BS-FWM conversion efficiency, due to the halving of the coupling coefficient for the fields \cite{McKinstrie4}. 

\textcolor{Black}{As far as the dominant physical mechanism which is responsible for noise in our experiments, we observe first that
for each pump k the product $\gamma P_k L\sim 0.15$. Following Ref. \cite{agraq}, and for the range of 
frequency detunings of our experiments, we can expect that at room temperature the Raman noise is significantly larger than the noise originating from spontaneous FWM, which moreover requires phase matching.}
Since the frequency conversion processes take place for relatively small detunings with respect to the Raman gain bandwidth, two important issues have to be considered when studying the noise figure \cite{McKinstrie4}. On one hand, the Raman gain in silica glass fibers has a relevant anisotropic contribution \cite{Lin8}, and on the other hand, spontaneous Raman noise at both Stokes and anti Stokes wavelengths exists at room temperatures \cite{Brainis8}. In particular, anti Stokes emission is possible due to distributed coupling with phonons. Indeed, for small sideband detunings, the phonons energy is comparable to $k_BT$, where $k_B$ is the Boltzmann constant and T is the temperature. Following Refs. \cite{Li6,Brainis8}, we analysed experimentally the spontaneous Raman emission which is separately induced by each of the two pumps in the absence of the signal. We then calculated the parallel ($R_{g,\parallel}$) and perpendicular ($R_{g,\perp}$) frequency-dependent Raman gain taking into account the input photon counts. In the right panel of Fig.\ref{Fig3} we present our experimental results obtained with the ECL laser at 1540.7nm (as P1) with 1.1 mW at the HNLF input. Similar results were obtained when using the IFSFLs pumps, but thanks to the narrow spectral width of the ECL, we could explore smaller frequency detunings. For each measurement, we averaged the number of detected photons over 60 seconds. The solid curves in Fig.\ref{Fig3} (right panel) provide the Raman model depicted in Ref. \cite{Lin8} for describing the parallel and perpendicular components of the Raman gain. That model includes both the isotropic ($R_{g,a}$) and anisotropic ($R_{g,b}$) contributions to the Raman gain. We found that this model fits our measurements very well, except for its perpendicular component on the Stokes side, where we observed a slightly larger orthogonal gain. Fig.\ref{Fig3} (right panel) shows that the anisotropic Raman contribution is relevant in our experiments, since for small detunings from the pumps this contribution is higher than the isotropic one \cite{Lin8}. As we have obviously $R_{g,\parallel} = R_{g,a}+R_{g,b}$, while $R_{g,\perp}= R_{g,b}/2$, the gain depolarization factor is therefore as high as 
$R_{g,\perp}/ R_{g,\parallel} \sim 0.3$. This is the reason why Raman noise was observed even in case C. Such a factor is expected to drop significantly when the signal detuning grows larger, owing to the predominance of the isotropic contribution.
\textcolor{Black}{Spontaneous FWM may explain some residual disagreements between the experimental results and the pure Raman noise assumption.}
 
\section{Conclusion}
To conclude, we explored the BS-FWM process and related polarization dependence in both strong signal and photon-counting regimes for small (20 nm) detunings from the pumps at telecom wavelengths. We also analysed its competition with noise, mainly of Raman origin. We observed a clear enhancement of the mean number of photons per second at the idler channel (1526 nm) when BS FWM was active, and the corresponding depletion of signal. We reported a significant difference of BS-idler counts between case A, where signal and pumps are all co-polarized and case C, where signal is cross polarized with co-polarized pumps (or case B, where signal is cross-polarized with pump 1 and cross-polarized with pump 2), which is in reasonable agreement with the expected 6 dB difference of the conversion efficiency. \textcolor{Black}{One may improve the conversion efficiency with a better suited fiber dispersion values and set of wavelengths; the idler bandwidth can be reduced by using an appropriate phase modulation for pumps  \cite{180nm}. Then, to go towards the single photon regime, smarter solution of detection based on time and spectral filtering should be implemented. Differently from high index contrast waveguides, HNLF fibers have naturally a ZDW close to telecom bandwidth but their full exploitation will require new generations of HNLF with specific control of the dispersion fluctuations along the fiber length as described in Ref.\cite{radi}.}

\section*{Acknowledgments}
We acknowledge the financial support from the Region Limousin and from the French National Research Agency, under grant ANR 08-JCJC-0122 PARADHOQS.

\end{document}